\documentclass[%
 reprint,
 superscriptaddress,
 amsmath,amssymb,
 aps,
]{revtex4-2}

\usepackage{graphicx}
\usepackage{dcolumn}
\usepackage{bm}
\usepackage{hyperref}

\graphicspath{{images/}}

\begin{document}

\preprint{APS/123-QED}

\title{Swift sky localization of gravitational waves\\ using deep learning seeded importance sampling}

\author{Alex Kolmus}
 \email{alex.kolmus@ru.nl}
 \affiliation{Institute for Computing and Information Sciences (ICIS), Radboud University Nijmegen, Toernooiveld 212, 6525 EC Nijmegen, The Netherlands}
\author{Grégory Baltus}
 \affiliation{STAR Institut, B\^{a}timent B5, Universit\'{e} de Li\`{e}ge, Sart Tilman B4000 Li\`{e}ge, Belgium}
\author{Justin Janquart}
 \affiliation{Nikhef, Science Park 105, 1098 XG Amsterdam, The Netherlands}%
 \affiliation{Institute for Gravitational and Subatomic Physics (GRASP), Utrecht University, Princetonplein 1, 3584 CC Utrecht, The Netherlands}
\author{Twan van Laarhoven}
 \affiliation{Institute for Computing and Information Sciences (ICIS), Radboud University Nijmegen, Toernooiveld 212, 6525 EC Nijmegen, The Netherlands}
\author{Sarah Caudill}
 \affiliation{Nikhef, Science Park 105, 1098 XG Amsterdam, The Netherlands}%
 \affiliation{Institute for Gravitational and Subatomic Physics (GRASP), Utrecht University, Princetonplein 1, 3584 CC Utrecht, The Netherlands}
 \author{Tom Heskes}
 \affiliation{Institute for Computing and Information Sciences (ICIS), Radboud University Nijmegen, Toernooiveld 212, 6525 EC Nijmegen, The Netherlands}
\date{\today}

\begin{abstract}
Fast, highly accurate, and reliable inference of the sky origin of gravitational waves would enable real-time multi-messenger astronomy. Current Bayesian inference methodologies, although highly accurate and reliable, are slow. Deep learning models have shown themselves to be accurate and extremely fast for inference tasks on gravitational waves, but their output is inherently questionable due to the blackbox nature of neural networks. In this work, we join Bayesian inference and deep learning by applying importance sampling on an approximate posterior generated by a multi-headed convolutional neural network. The neural network parametrizes Von Mises-Fisher and Gaussian distributions for the sky coordinates and two masses for given simulated gravitational wave injections in the LIGO and Virgo detectors. We generate skymaps for unseen gravitational-wave events that highly resemble predictions generated using Bayesian inference in a few minutes. Furthermore, we can detect poor predictions from the neural network, and quickly flag them. 
\end{abstract}

\maketitle

\section{\label{sec:introduction}Introduction}
Gravitational waves (GWs) have immensely advanced our understanding of physics and astronomy since 2015 \cite{abbott2016tests, abbott2017gw170814, yamada2019testing, baiotti2019gravitational}. These GWs are observed by the Hanford (H) and Livingston (L) interferometers of the Laser Interferometer Gravitational Wave Observatory (LIGO) \cite{aasi2015advanced} and the Advanced Virgo (V) interferometer \cite{acernese2014advanced}. The collaboration between these three detectors has enabled triple-detector observations of GWs \cite{abbott2017gw170814}, making it possible to do proper sky localisation of their astrophysical sources. This additional detector changes the sky distribution from a broad band to a more narrow distribution \cite{abbott2017gw170814}.

Better early sky localisation capabilities would allow for real-time multi-messenger astronomy (MMA), observing astrophysical events through multiple channels - electromagnetic transients, cosmic rays, neutrinos - only seconds after the GW is detected. MMA is limited to GWs originating from binary neutron star (BNS) and neutron star-black hole mergers. According to current literature, it is unlikely that binary black holes (BBHs) emit an electromagnetic counterpart during their merger \cite{doctor2019search, perna2019limits}. Currently, astrophysicists try to collect the non-GW channels in the weeks after the event. A notable example is GW170817 \cite{abbott2017gw170817, cowperthwaite2017electromagnetic}. This process takes an enormous amount of effort, while the obtained data quality is often sub-optimal. Having all channels observed for the full duration of the event would be a major leap forward. Real-time MMA would enable a plethora of new science, e.g. unravelling the nucleosynthesis of heavy elements using r- and s-processes, more accurate and novel tests of general relativity, and a deeper understanding of the cosmological evolution \cite{ berti2018extreme, barack2019black, fishbach2019standard}. As aforementioned, real-time MMA relies on the generation of a skymap and it imposes two limits on the methodology used to obtain one. First, it needs to be swift in order to allow observatories to turn towards an event's origin, preferably only seconds after its observation. Second, the skymap needs to be as accurate as possible since telescopes have a limited area they can observe. Below we present current approaches in generating skymaps for GW events.

Most GW software libraries \cite{veitch2015parameter, biwer2019pycbc} use Bayesian inference methods - in particular Markov chain Monte Carlo (MCMC) and nested sampling \cite{skilling2006nested} - to construct the posterior over all GW parameters. These methods asymptotically approach the true distribution given a sufficient number of samples \cite{murphy2012machine}. Although theoretically optimal, a chain with around $10^6$ to $10^8$ samples is required \cite{veitch2015parameter} to closely approximate the true posterior distribution for a GW event. Even when using Bilby \cite{ashton2019bilby} - a modern Bayesian inference library made for GW astronomy - to perform the inference for a single BBH event, takes hours to produce \cite{gabbard2019bayesian}; BNS events take even longer. Bayesian inference is the most accurate method available for GW posterior estimation, but its run-time is prohibitively long when it comes to MMA.

To overcome the speed limitations of the Bayesian approaches, Singer and Price developed BAYESTAR in 2016 \cite{singer2016rapid}, an algorithm that can output a robust skymap for a GW event within a minute. BAYESTAR realizes this speedup in two ways. First, it exploits the information provided by the matched filtering pipeline used in the detection of GWs. The inner product between time strain and matched filters contains nearly all of the information regarding arrival times, amplitudes and phases, which are critical for skymap estimation. Second, Singer and Price derive a likelihood function that is semi-independent from the mass estimation and does not rely on direct computation of GW waveforms, allowing for massive speedups and parallelization. Although BAYESTAR is fast, its predictions tend to be broader and less precise than those made by Bilby \footnote{The GWTC-2 catalog \cite{abbott2021gwtc} data release provides skymaps made using Bayesian inference methods for recent events. Comparison with the skymaps made by BAYESTAR can be made by looking at skymaps on https://gracedb.ligo.org/latest/.}.

Deep learning (DL) algorithms have shown themselves to be exceptionally quick and powerful when handling high-dimensional data \cite{lecun2015deep, schmidhuber2015deep}. Therefore, they are an interesting alternative to the Bayesian methods. Several papers have proposed methods to estimate the GW posterior, including the skymap, using DL algorithms. Examples of such algorithms are Delaunoy et al. \cite{delaunoy2020lightning} and Green and Gair \cite{green2021complete}. Delaunoy et al. \cite{delaunoy2020lightning} use a convolutional neural network (CNN) to model the likelihood-to-evidence ratio when given a strain-parameter pair. By evaluating a large amount of parameter options in parallel, they can generate confidence intervals within a minute. The reported confidence intervals are slightly wider than those made by Bilby. A completely different approach was taken by Green and Gair \cite{green2021complete}. They showcase complete 15-parameter inference for GW150914 using normalizing flows. They apply a sequence of invertible functions to transform an elementary distribution into a complex distribution \cite{papamakarios2021normalizing} which, in this case, is a BBH posterior. Within a single second, their method is able to generate 5,000 independent posterior samples that are in agreement with the reference posterior\footnote{Throughout this paper, reference posterior is used to imply a posterior that is generated using Bayesian inference.}. A Kolmogorov-Smirnov test confirms that these samples are very closely resemble the samples that are drawn from the exact posterior. Both DL methods are fast and seem to be accurate for the 100 - 1000 simulated GW events they have been evaluated on. However, these methods have a few issues: (1) they are both susceptible to changes in the power spectral density (PSD) and signal-to-noise ratio (SNR), (2) both are close in performance to Bilby but do not match it, (3) they can act unpredictably outside of the trained strain-parameters pairs and, even within this space, they can act unpredictably due to the blackbox nature of neural networks (NNs). Issues (1) and (2) have been addressed for the normalizing flow algorithm in a recent paper by Dax et al. \cite{dax2021real}, however the robustness guarantees remain behind those of traditional Bayesian inference.

Our method tries to bridge the gap between Bayesian inference and DL methods, allowing for fast inference while still guaranteeing optimal accuracy. It is to be noted that combining Bayesian inference and DL methods has recently gained traction in the GW community, see for example reference \cite{williams2021nested}. The goal of our algorithm is to restrict the parameter space such that, via sampling, one can quickly obtain an accurate skymap. We use a multi-headed CNN to parameterize an independent sky and mass distribution for a given BBH event. The model is trained on simulated precessing quasi-circular BBH signals resembling the ones observed by the HLV detectors. The parameterized sky and mass distributions are Gaussian-like and are assumed to approximate the sky and mass distributions generated by Bayesian inference. Using the parameterized sky and mass distributions, we construct a proposal posterior in which all other BBH parameters are uniformly distributed. By using importance sampling we can then sample from the exact reference posterior. This implies that we effectively match the performance of Bayesian inference in a short time span, without exploring the entire parameter space. We stress that this work is a proof of concept to show the promises of combining NNs and Bayesian inference. More flexible DL models and BNS events will be considered in future studies.

This paper is organised as follows. Section 2 discusses the model architecture and importance sampling scheme. Section 3 details the performed experiments, including the model training. Section 4 covers the results of these experiments and subsequently assesses the performance of the model and importance sampling scheme by comparing it with skymaps generated using Bilby for a non-spinning BBH system. Conclusions and future endeavours are specified in Section 5.

\section{\label{sec:methodology}Methodology}
Our inference setup is a two-step method. In the initial step we infer simple distributions for the sky localization and the masses of the BBH by using a neural network. Subsequently, we apply importance sampling to these simple distributions to compute a more accurate posterior. The first subsection describes the role and implementation of importance sampling. The second subsection discusses the neural network setup and our method for distribution estimation. 

\subsection{Importance sampling}
High-dimensional distributions in which the majority of the probability density is confined to a small volume of the entire space are hard to sample from, which results in long run times to get proper estimates when using MCMC methods. A well-known method to cope with this problem is importance sampling. By using a proposal distribution $q$ that covers this high probability density region of the complex distribution $p$ one can quickly obtain useful samples. There are two requirements when using importance sampling. First, the desired distribution $p$ needs to be known up to the normalization constant $Z$: $p(\lambda)=\frac{1}{Z}\theta(\lambda)$. Second, the proposal distribution $q$ needs to be non-zero for all $\lambda$ where $p$ is non-zero. Importance sampling can be understood as compensating for the difference between the distributions $p$ and $q$ by assigning an importance weight $w(\lambda)$ to the each sample $\lambda$,
\begin{equation}
    w(\lambda) = \frac{\theta(\lambda)}{q(\lambda)},
\end{equation}
where the fraction is the likelihood ratio between the - not-normalized - $p$ and $q$. The distribution created by the reweighted samples will converge to the $p$ distribution given enough samples \cite{bishop2006machine}.

Generating accurate posteriors for GW observations using MCMC is very time consuming, and thus importance sampling is an interesting alternative. Importance sampling requires us to have a viable proposal distribution. Published posteriors for known gravitational waves show that the probability density in the posterior is relatively well confined for both the sky location and the two masses \cite{abbott2021gwtc}. A Von Mises Fisher (VMF) and Multi Variate Gaussian (MVG) distribution are good first order approximations of the sky and mass distribution respectively, and thus suitable to use as a proposal distribution for importance sampling. We propose to construct this proposal distribution by assuming a uniform distribution over all non-spinning BBH parameters, except for the sky angles which will be represented by a VMF and a MVG distribution for the masses. Assuming that the BBH parameters, sky angles, and masses are independent, our proposal distribution becomes the product of these two distributions. In the next subsection we discuss how we create this proposal distribution using a neural network.

Importance sampling demands a likelihood function for the proposal distribution and the desired distribution. In the previous paragraph we have discussed how we want to create a proposal distribution, we will now focus on the desired distribution $p$. For the likelihood function of the GW posterior $p(s|\lambda)$ we take the definition given by Canizares et al. \cite{canizares2013gravitational}:
\begin{equation}
    p(s|\lambda) \propto \theta(s|\lambda) = \exp\left(-\frac{\langle s - h(\lambda)| s - h(\lambda)\rangle}{2} \right),
\end{equation}
where $s$ is the observed strain, $h(\lambda)$ is the GW template defined by parameters $\lambda$. The inner product is weighted by the PSD of the detector's noise. In practice we use the likelihood implementation provided by Bilby named \textit{GravitationalWaveTransient}.

We now have all the parts needed to discuss how we utilize importance sampling for a given strain $s$. A trained neural network parameterizes the proposal distribution $q$ for the given strain. The proposal distribution generates $n$ samples, these samples represent possible GW parameter configurations. For each sample we calculate the logarithm of the importance weight,
\begin{equation}
    \log w(\lambda) = \log \theta(s|\lambda) - \log q(\lambda) + C,
\end{equation}
instead of the importance weight $w(\lambda)$ itself to prevent numeric under- and overflow. The constant $C$ is added to set the highest $\log w(\lambda)$ to zero, to prevent very large negative values from becoming zero when we calculate the associated likelihood. Since we normalize the weights afterwards the correct importance weights are still obtained. The reweighted samples represent the desired distribution $p$.

If the proposal distribution does not cover the true distribution well enough, the importance samples will be dominated by only a single to a few weights if we restrict the run-time. We can use this as a gauge to check if the skymap produced by the neural network and importance sampling is to be trusted.

\subsection{Model}
Previous work done by George et al. \cite{george2018deep} shows that convolutional neural networks (CNN) are able to extract the masses from a BBH event just as well as the currently-in-use matched filtering. Furthermore, work done by Fan et al. \cite{fan2019applying} indicates that 1D CNNs are able to locate GW origins. We therefore chose to use a 1D CNN to model both the distribution across the sky for the origin of the GWs and a multivariate normal distribution for the two masses of the BBH system. 

\begin{figure}
    \centering
    \includegraphics[width=\linewidth]{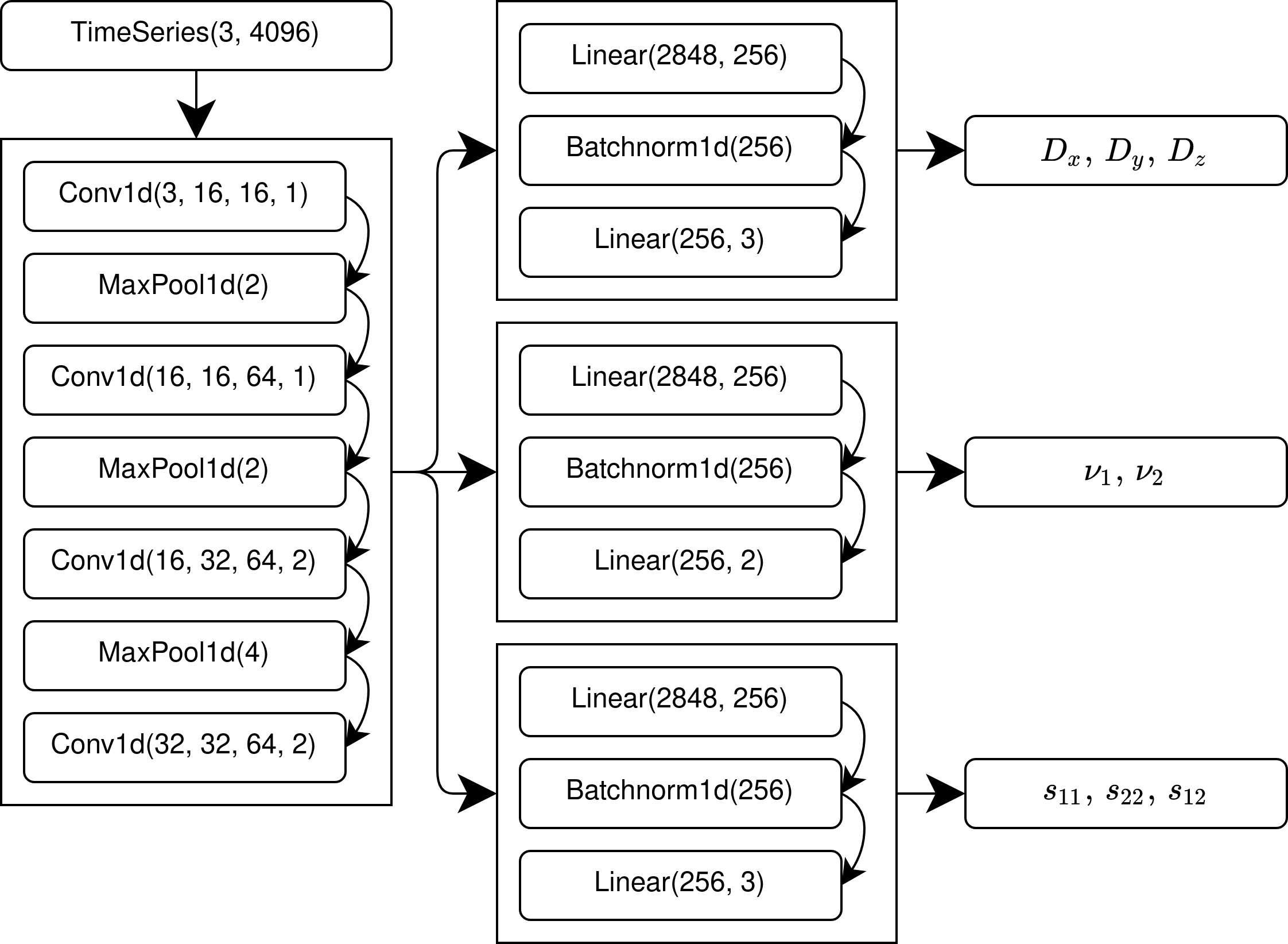}
    \caption{A graphical depiction of the convolutional neural network used in this work. After each MaxPool1d and Batchnorm1d layer a leaky ReLU activation function with an $\alpha=0.1$ is applied. The convolutional part is shown on the left and takes as input a time series of 4096 elements with 3 channels. Conv1D($i$, $o$, $k$ , $d$) denotes a 1D convolution with $i$ input channels, $o$ output channels, kernel size $k$ and dilation factor $d$. MaxPool1d($k$) denotes a 1D max pooling layer with kernel size $k$. The output of the convolutions is given to three independent neural network heads. The first head predicts the sky location parameterized as $D=(D_x, D_y, D_z)$, the second head predicts the mean of the masses of the two black holes, and the last head predicts the uncertainty elements of the covariance matrix over the two masses. Linear($i$, $o$) denotes a linear transformation with $i$ input features and $o$ output features. Lastly, Batchnorm1d($i$) denotes a 1D batch normalization layer with $i$ input features.}
    \label{fig:model}
\end{figure}

The network architecture of this 1D CNN is presented in Figure \ref{fig:model} and consists of four parts: a convolutional feature extractor and three neural network heads. These heads are used to specify the two distributions. The following properties were tested or tuned for optimal performance: number of convolutional layers, kernel size, dilation, batch normalization, and dropout. The model shown in Figure \ref{fig:model} produced the best result on a validation set. 

The convolutional feature extractor generates a set of features that characterize a given GW. This set of features is passed on to the neural heads. Each head is specialized to model a specific GW parameter. The first head determines the sky distribution, the second head the masses, and the third head the uncertainty over the two masses. Below we will elaborate on each of these heads and how they characterize these distributions.

The first head specifies the distribution of the GW origin. Since the sky is described by the surface of a 3D sphere, a 2D Gaussian distribution is an ill fit. A suitable alternative is the Von Mises-Fisher (VMF) distribution \cite{fisher1993statistical} which is the equivalent of a Gaussian distribution on the surface of a sphere. The probability density function and the associated negative log-likelihood (NLL) of the VMF distribution:
\begin{align}
    p(x|\mu, \kappa) &= \frac{\kappa}{4\pi \sinh(\kappa)} \exp\left(\kappa x^T\mu\right)\\
    NLL_{\text{VMF}}(x, \mu, \kappa) &= -\log(\kappa) - \log(1 - \exp(-2\kappa)) \nonumber\\- \kappa  - &\log(2\pi) + \kappa x^T\mu \: ,
\end{align}

where $x$ and $\mu$ are normalized vectors in $\mathbb{R}^3$, with the former being the true direction and the latter being the predicted direction. $\kappa$ is the concentration parameter, which determines the width of the distribution. It plays the same role as the inverse of the variance for a Gaussian distribution. We use this distribution by letting the first head output a three-dimensional vector $D=(D_x, D_y, D_z)$. The norm of $D$ specifies the concentration parameter $\kappa$, and its projection onto the unit sphere gives the mean $\mu$, $\kappa=|D|$, and $\mu=D/|D|$. These values together with the true direction $x$ are used to calculate the negative log-likelihood, which is used as the loss function of the first head.

The second and third neural heads specify a 2D multivariate Gaussian (MVG), which describes the possible configurations of the masses. The means $\nu$ of the MVG are given by the second head and the covariance matrix $\Sigma$ is specified by the third head. Given the true values of the masses $y=(m_1, m_2)$ the probability density function and associated negative log-likelihood of the MVG are:
\begin{align}
    p(y|\nu, \Sigma) &= \frac{1}{\sqrt{(2\pi)^2|\Sigma|}} \nonumber\\
    &\exp\left(-\frac{1}{2}(y - \nu)^T \Sigma(y-\nu) \right) \\
    NLL_{\text{MVG}}(y, \nu, \Sigma) &= \frac{1}{2}(y - \nu)^T\Sigma^{-1}(y-\nu) + \nonumber\\ &\frac{1}{2}\log\left(|\Sigma|\right) +  \log\left(2\pi\right)  \: .
\end{align}

The inverse covariance term in the negative log-likelihood can contain imaginary numbers if the covariance matrix is not positive-definite. To ensure that the covariance matrix $\Sigma$ remains positive-definite, it is parameterized through:
\begin{align}
    \Sigma_{11} &= \exp(s_{11}) \\
    \Sigma_{22} &= \exp(s_{22}) \\
    \Sigma_{21} &= \Sigma_{12} = \tanh(s_{12})\sqrt{\Sigma_{11}\Sigma_{22}} \: .
\end{align}

The three variables $s_{11},\, s_{22}, \, s_{12}$ are predicted by the third neural head and define the covariance matrix completing the MVG prediction of the masses. The paramerization and implementation of the MVG is based on the work of Russell et al. \cite{russell2021multivariate}. 

By further assuming that the sky distribution is independent of the mass distribution, we obtain a first approximation of the posterior distribution, thereby satisfying the requirements for importance sampling.

\section{\label{sec:experiments}Experiments}
Experiments were performed on two different fronts: (1) training the neural network followed by the empirical evaluation of its performances on unseen test data, and (2) comparing the neural network model, importance sampling scheme, and Bilby based on several metrics and skymaps. Below we describe the experimental details and justify decisions we made. All experiments were performed on a computer with a 16-core AMD Ryzen 5950X CPU, NVIDIA 3090 RTX GPU, and 64 GB of RAM. Source code is available at \url{https://github.com/akolmus/swiftsky}.  

\subsection{Training and evaluating the neural model}
To obtain strain-parameter pairs for training and validation, we sampled parameters from a BBH parameter prior (see Table \ref{table:data_prior}) and generated the associated waveforms using the \textit{IMRPhenomPv2} waveform model \cite{hannam2014simple}. The waveforms were generated in the frequency domain in the frequency band of 20 to 2048 Hz. The duration of the signal is 2 seconds. Subsequently, these waveforms were projected onto the HLV interferometers. We sampled the SNR from a scaled and shifted Beta distribution with its peak set to 15 (see Figure \ref{fig:snr_distribution}). The luminosity distance in the prior was set to a 1000 Mpc and scaled afterwards to match the desired SNR. We generated Gaussian noise from the design sensitivity PSD for each detector. Finally, the signal was injected into the noise and an inverse Fourier transform was applied to obtain the strains as time series. This setup allowed us to generate an arbitrary amount of unique strain-parameter pairs, which resulted in every training epoch having a unique dataset.

We applied three preprocessing steps to the data. All time series were whitened with the aforementioned PSDs. Next, the time series were normalized. A normalizer was calculated such that noise-only strains have mean zero and a standard deviation of one. We found empirically that calculating a normalizer for the noise instead of noise plus signal allowed the neural network to converge faster and achieve lower losses. Lastly, to make the mass distribution easier to learn we calculated a shift and scaling factor for the target masses such that all target masses were between -1 and +1. The shifting and scaling were applied inversely to the neural network output during importance sampling to get the correct masses. 

The model was trained for 300 epochs with a batch size of 128. During each epoch we drew 500\,000 strain-parameter pairs for training and 100\,000 strain-parameter pairs for validation. Adam \cite{kingma2015adam} was used to optimize the model in conjunction with a cosine annealing scheme with warm restarts \cite{Loshchilov2017sgdr}. The learning rate oscillated between $10^{-3}$ and $10^{-5}$ with a period of 20 epochs; weight decay was set to $10^{-6}$. Multiple hyperparameter configurations were tested; this configuration obtained the best performance.

In order to benchmark the trained model, an unseen test set was generated of 100\,000 strain-parameter pairs at specific SNR values. The model was evaluated using the mean absolute angular error (maae) and the average 90\% confidence area of the predicted VMF distributions.

\subsection{Applying and evaluating importance sampling}
To evaluate the importance sampling procedure, we constructed a slightly simpler test set in which we restricted the maximum spin magnitude to be zero. This was done to limit the Bilby run-time. The importance sampling procedure discussed in Section 2.2 was applied to the first 100 strain-parameter pairs of this test set at three different optimal SNR values: 10, 15, and 20. For each strain-parameter pair we generated 200\,000 importance samples. In order to simulate multiple independent runs at various time points for the same strain-parameter pair, we subsampled from these 200\,000 importance samples during the experiments. 

We ran two experiments to test the convergence of the importance sampling method. In the first experiment, we used the importance sampling scheme as a maximum likelihood estimator. For a given set of importance samples we chose the sample with the highest likelihood and calculated the angle between this sample and the true sky coordinates. In the second experiment, we represented the probability density function of the importance samples by a kernel density estimator and tested how well the resulting density covered the true right ascension. Specifically, we used a Gaussian kernel density estimator\footnote{The \textit{gaussian\_kde} from the \textit{scipy} python package.} to fit the right ascension distribution proposed by the importance samples. The log-likelihood of the actual right ascension was used to measure the quality of the estimated density. We removed a few outliers from the second experiment, by restricting ourselves to only the right ascension the number of outliers was reduced. These outliers had densities that did not cover the true right ascension at all, resulting in extreme negative log-likelihoods which dominate the average log-likelihood. For both experiments we expect the metric to improve as the number of importance samples increases, and to level after a significant number of importance samples indicating convergence. 

\subsection{Generating skymaps}
We use Bilby as a benchmark to generate skymaps for the first ten strain-parameter pairs of the test set and for each create a version at an SNR of 10, 15, and 20. To make a fair comparison, the prior given to the Bilby sampler has its spin components set to zero. Moreover, the posterior inference was performed with standard settings, and each run took between 2.5 and 7 hours to complete. During these runs the live points of the sampler were saved every 5 seconds and labelled by the total number of sampled points. These saved points were used to run the two importance sampling experiments for Bilby. 

\section{\label{sec:results}Results}
\subsection{CNN}
\begin{figure}
    \includegraphics[width=\linewidth]{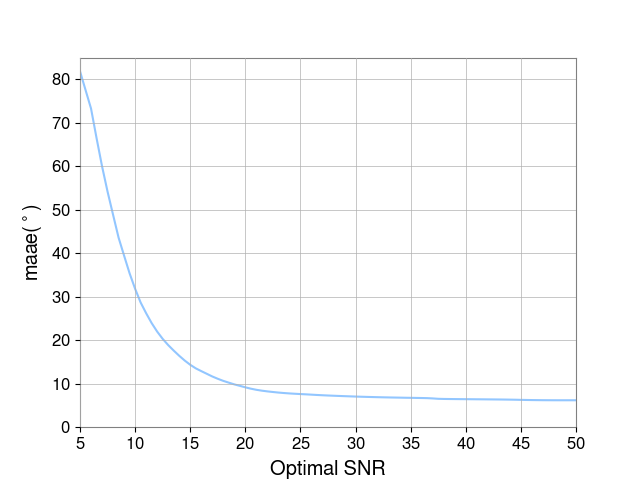}
    \includegraphics[width=\linewidth]{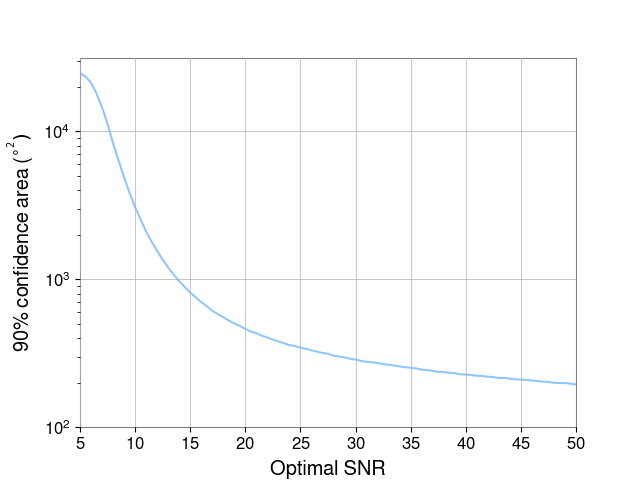}
    \caption{Characterization of the neural network in terms of accuracy and certainty over the test. Left: the maae (mean absolute angular error) between the sky angle predicted by the model and the actual sky location as a function of the SNR. Right: the average size of the 90\% confidence area, expressed in degrees squared, of the predicted VMF distributions as a function of the SNR.}
    \label{fig:nn_results}
\end{figure}

In Figure \ref{fig:nn_results} we summarize the results for the first experiment: the left panel gives the mean absolute angular error (maae) in the sky location  and in the right panel we plot the 90\% confidence area of the VMF distribution. As expected, as the SNR increases the prediction error in the sky location decreases and the 90\% confidence area becomes smaller. The error in the mass prediction is similar to those of other CNN approaches \cite{george2018deep}, see Figure \ref{fig:nn_mass}, indicating that the setup works well. We do note that the error in the sky location seems to be quite high for $\text{SNR}<10$ and that it does not converge to zero for high SNR. We can think of two possible explanations for the poor performance at low SNR. First, the detection rate using either CNNs or matched filtering pipelines at an SNR of 5 is less than 40\% \cite{george2018deep, gebhard2019convolutional}. At such a low SNR, it is difficult for the model to discern the differences in arrival time at each detector, which explains the slightly better than random predictions for $\text{SNR}<7$. When we compare our angular error with other CNN approaches \cite{chatterjee2019using, fan2019applying}, the average error seems to be similar. Furthermore, Chua and Vallisneri \cite{chua2020learning} reported that Gaussian approximations are only accurate for high SNR ($\text{SNR } > 8$) and even then multimodality might arise. Second, the sky distribution can be multimodal. This multi-modality is either due to strong noise or can be due to a sky reflection \cite{veitch2015parameter}. For three detectors, there are two viable solutions to the triangulation problem: the true sky location and its reflection. In most cases the amplitude information is sufficient to break the degeneracy between the location and its reflection. However, at certain sky angles this amplitude information does not lift the degeneracy and a multimodal distribution is required. For these angles the model has a 50\% chance of guessing the wrong mode and thus having an average angular error of 90$^\circ$. 

\subsection{Importance sampling}
\begin{figure}
    \includegraphics[width=\linewidth]{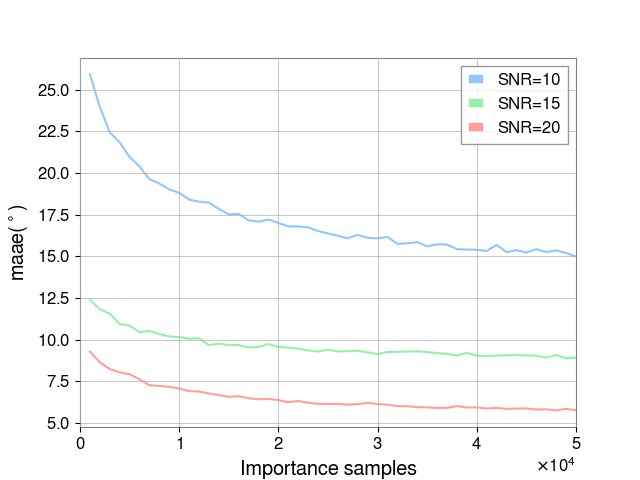}
    \includegraphics[width=\linewidth]{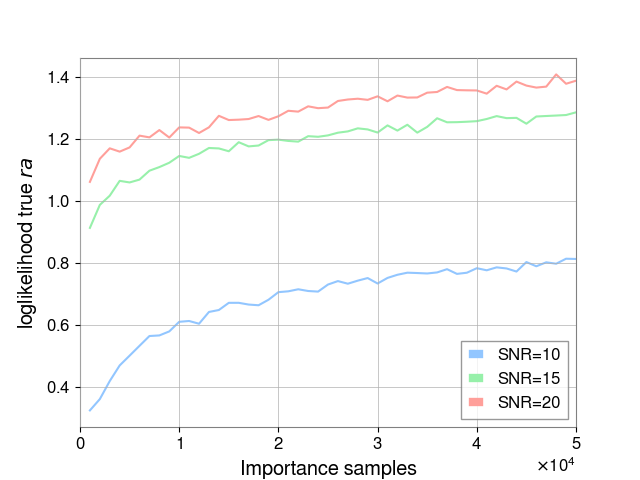}
    \caption{Characterization of the importance sampling, with the number of importance samples ranging from a 1\,000 to 50\,000. The colors represent different SNR values with blue, green, and red being 10, 15, and 20 respectively. Left: the maae of the importance sample with the highest likelihood as a function of the sample size. Right: the log-likelihood of the true right ascension according to the kernel density estimator created by importance samples as a function of sample size.}
    \label{fig:imp_results}
\end{figure}

The results of the importance sampling experiments are shown in Figure \ref{fig:imp_results}.  The left panel shows the maae as the number of importance samples increases. The right panel shows the log-likelihood of the true right ascension given by kernel density based on a varying number of importance samples. The majority of the convergence in the maae seems to happen within the first 30\,000 samples. The slow convergence can largely be attributed to strains for which the model predicted a wide sky distribution. When we compare this to results of Bilby, we see that the maae of the highest likelihood sample for all SNR is always between 1 and 8 degrees. Importance sampling is competitive for an SNR of 20 and is close for an SNR of 15, especially when we consider that in both cases 2 out of the 100 sky distributions were parameterized as the sky reflection.
However, importance sampling is not competitive with Bilby in the second experiment. For all SNR values Bilby reports log-likelihoods between 2 and 3, see Figure \ref{fig:bilby_likelihood}, and importance sampling does not reach these values. If we consider runs that show good convergence, i.e. where 90\% of the importance weight is not determined by less than ten importance samples, importance sampling also reports log-likelihoods between 2 and 3. In Figure \ref{fig:imp_redone} we have repeated the kernel density experiment, but only for the well-converged runs. These runs represent 30\% of all runs, and almost no $\text{SNR}<10$ runs. 
\begin{figure*}
    \includegraphics[width=\linewidth]{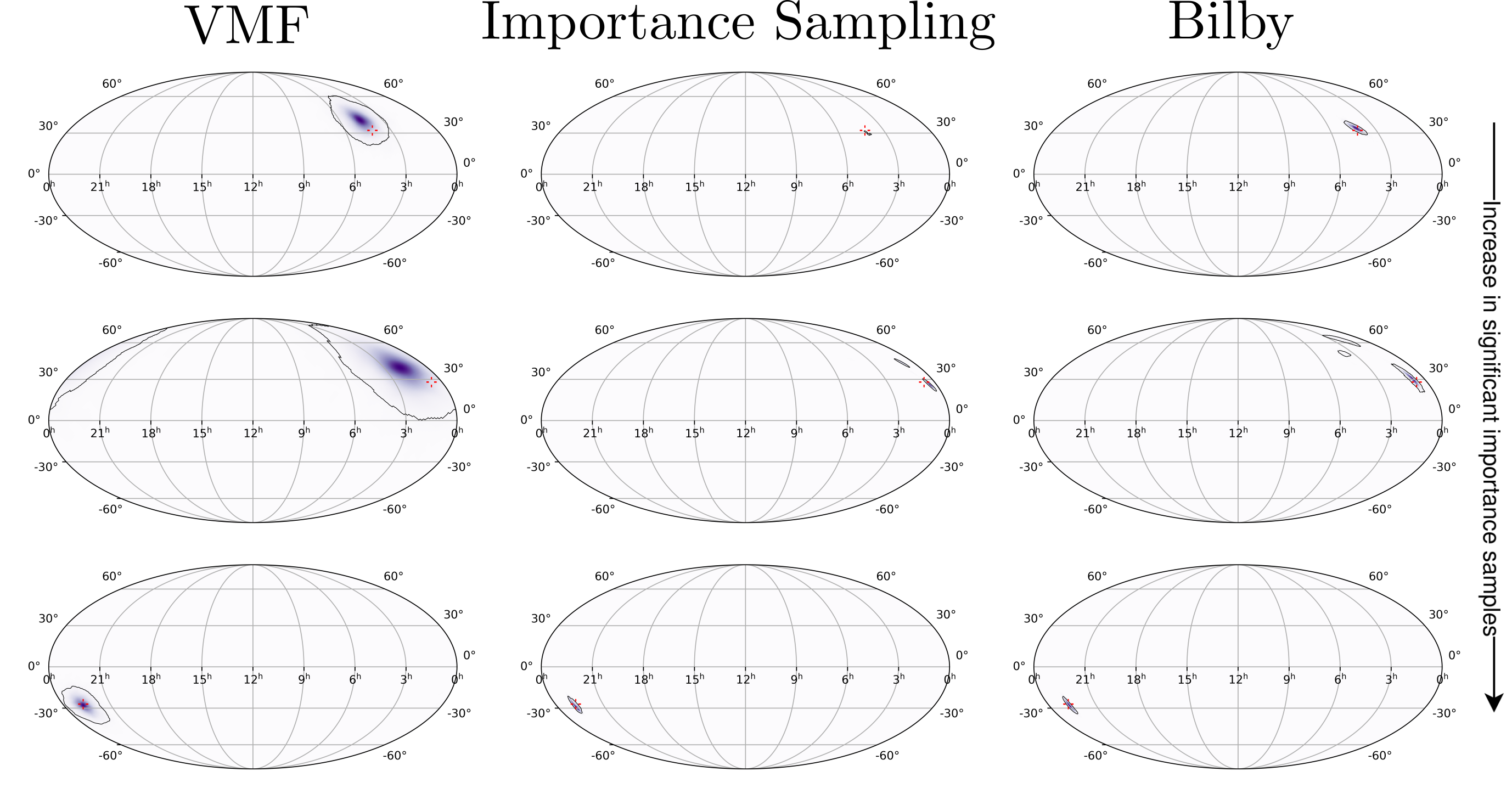}
    \caption{Examples of predicted skymaps by our neural network (left), importance sampling after 100\,000 steps or roughly 5 minutes of computing time (middle), Bilby at convergence (right). The Bilby runs took at least 3 hours to complete. The true sky location is indicated in red. The shown skymaps were generated for signals with an SNR of 15. The number of significant importance samples, and hence the quality of the sky maps, increases as we go from the top row to the bottom row.}
    \label{fig:imp_skymaps}
\end{figure*}

\subsection{Generating skymaps}
As a final test, we generated skymaps using the neural network, importance sampling, and Bilby on the same signals. Three representative skymaps are shown in Figure \ref{fig:imp_skymaps}. The skymaps generated by the neural network are significantly more spread out than those generated by importance sampling and Bilby. As we explained in the previous sections, this might due to the neural network overestimating the uncertainty and having difficulty extracting the exact signal from the detector noise. The skymaps generated by importance sampling and Bilby resemble each other quite a lot, their peak intensities are in the same position and the sky distributions occupy roughly the same area. However, the importance sampling skymaps are grainy and sometimes do not cover the complete area that Bilby does. As can be seen in the bottom row of Figure \ref{fig:imp_skymaps}, when the predicted VMF distribution has its peak intensity on the correct position the importance sampling creates better looking sky maps. This improvement is due to the increased number of significant importance samples. These results indicate that a larger number of significant importance samples is needed, which is to be expected with only 5 minutes of run-time. Within only 1-4\% of the Bilby run-time we are already able to recover the essentials of the skymaps. \bigskip

\section{\label{sec:conclusion}Conclusion}
In this paper, we produced skymaps for simulated BBH events using an importance sampling scheme that turns an approximate skymap made by a neural network into a skymap that represents the exact Bayesian posterior distribution. Experiments show that our method is competitive with Bilby and can produce the essentials of the skymap within 4\% of the Bilby run-time. However, in some cases the proposal distributions made by the neural network are too crude, which hampers the efficiency of the importance sampling scheme. If the sampling efficiency is improved further, importance sampling could be used as a quick alternative to Bilby or LALInference for inferring the GW posterior. In future work, we will consider more advanced deep learning models such as normalizing flows to infer more accurate posterior distributions.

\begin{acknowledgments}
This work was (partially) funded by the NWO under the CORTEX project (NWA.1160.18.316). G.B. is supported by aFRIA grant from the Fonds de la Recherche Scientifique-FNRS, Belgium. J.J. is supported by the research program of the Netherlands Organisation for Scientific Research (NWO).
\end{acknowledgments}

\bibliography{manuscript}

\onecolumngrid
\appendix

\section{Training details}
\begin{figure}[ht]
    \includegraphics[width=0.6\linewidth]{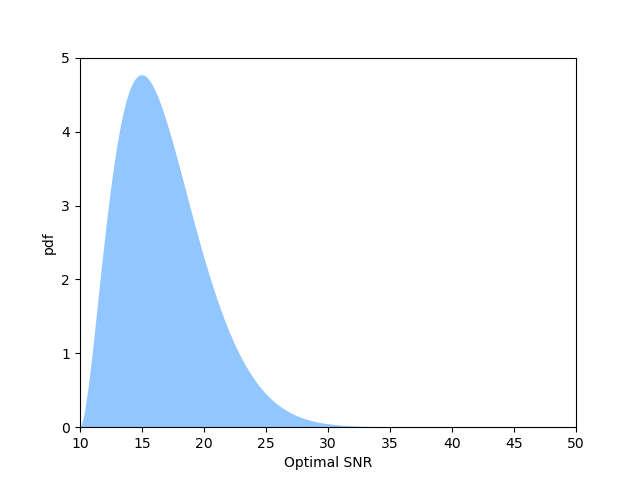}
    \caption{Scaled and shifted Beta distribution that acts as the SNR sampling distribution during training and validation. The vertical axis represents the probability density function of this Beta distribution, the horizontal axis represents the SNR value.}
    \label{fig:snr_distribution}
\end{figure}

Here we show the priors used for data generation (see Table \ref{table:data_prior}) and the SNR distribution during training (see Figure \ref{fig:snr_distribution}). 
\begin{table}[ht]
    \caption{The priors used for the data generation. The luminosity distance in the prior was set to a 1000 Mpc and scaled afterwards to match the desired SNR.}
    \label{table:data_prior}
    \begin{ruledtabular}
        \begin{tabular}{lcccc}
        parameter                & prior & minimum & maximum & unit \\ \hline
        Masses (constraint)      & - & 20      & 80      & $\textup{M}_\odot$ \\ 
        Chirp mass               & Uniform & 10      & 100     & $\textup{M}_\odot$ \\
        Mass ratio               & Uniform & 0.25    & 1.0     & -    \\
        Spin magnitudes          & Uniform & 0       & 0.95    & -    \\ 
        Spin polar angles        & Sine & 0       & $\pi$   & rad  \\ 
        Spin azimutal angles     & Uniform & 0       & 2$\pi$  & rad  \\ 
        Right ascension          & Uniform & 0       & 2$\pi$  & rad  \\ 
        Declination              & Cosine & -0.5$\pi$& 0.5$\pi$& rad  \\ 
        Binary inclination angle & Sine & 0       & $\pi$   & rad  \\ 
        Coalescence phase angle  & Uniform & 0       & 2$\pi$  & rad  \\ 
        Polarization angle       & Uniform & 0       & 2$\pi$  & rad  \\ 
        Time Shift               & Uniform & -0.1    & 0.1     & s    \\ 
        Luminosity distance      & - & 1000    & 1000    & Mpc  \\ 
        \end{tabular}
    \end{ruledtabular}
\end{table}

\section{Mass estimation performance}
\begin{figure}[hb]
    \centering
    \includegraphics[width=0.6\linewidth]{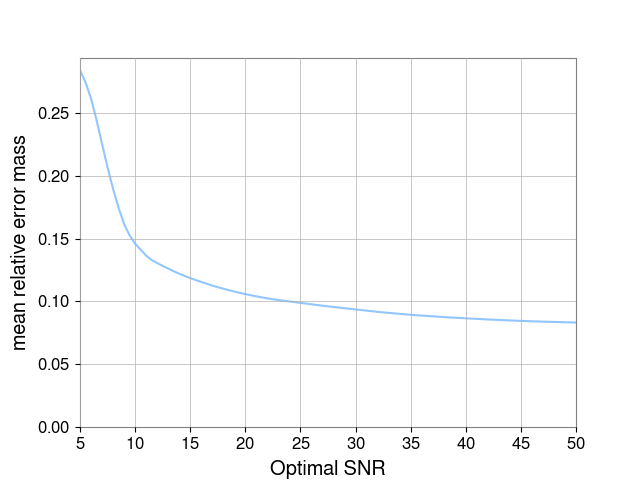}
    \caption{The mean relative error of the estimated masses by the neural network on the test set as a function of the optimal SNR. It is almost identical to the Figure 5 in \cite{george2018deep}.}
    \label{fig:nn_mass}
\end{figure}

In Figure \ref{fig:nn_mass}, we show the mean relative error of the estimated masses over the test set. This figure closely resembles Figure 5 in \cite{george2018deep}. Any differences are due to the difference in setup. The main differences are that our priors include spins and that we do not use a stationary sky origin.

\section{Importance sampling}
We redid the importance experiment with only well converged runs, the shown log-likelihood values are close to those reported by Bilby.
\begin{figure}
    \centering
    \includegraphics[width=0.6\linewidth]{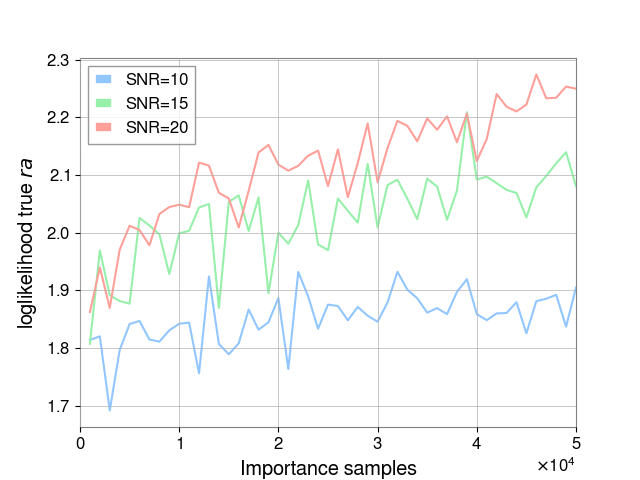}
    \caption{The log-likelihood of the true right ascension according to the kernel density estimator using only the importance samples of well converged runs. These values are more in line with those of Bilby, see Appendix D.}
    \label{fig:imp_redone}
\end{figure}

\section{Bilby run}
For thirty Bilby runs, ten per SNR value, we repeated the experiments reported in Section 3.2. Below we show the results for one of the ten samples.

\begin{figure}[ht]
    \centering
    \includegraphics[width=0.6\linewidth]{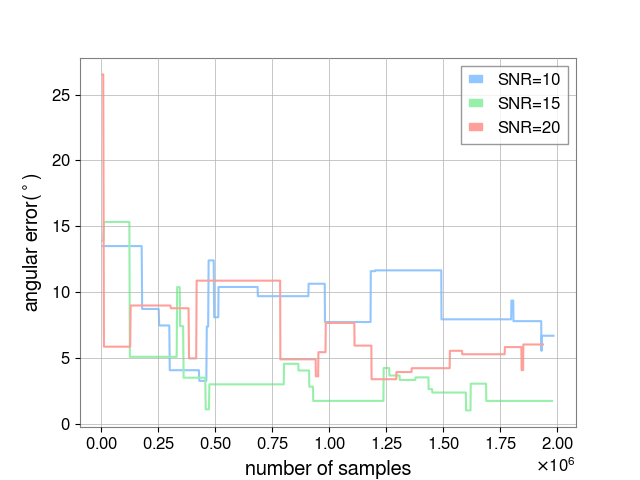}
    \caption{The angle between the sky location of the highest likelihood sample and the actual sky location. The vertical axis represents how many samples Bilby has generated (live and dead samples).}
    \label{fig:bilby_angles}
\end{figure}

\begin{figure}[hb]
    \centering
    \includegraphics[width=0.6\linewidth]{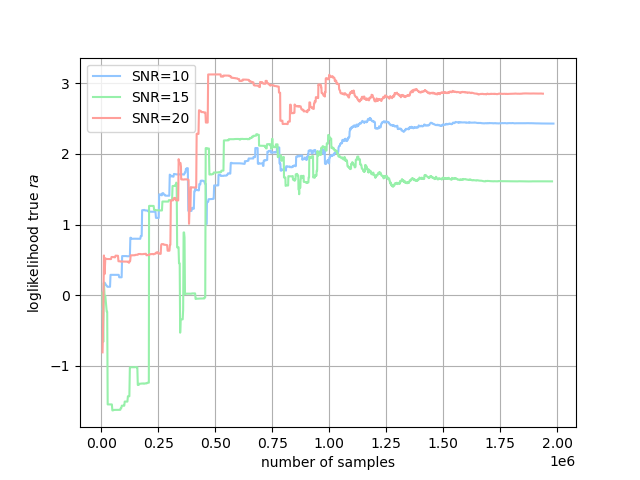}
    \caption{The loglikelihood of the true right ascension according to the kernel density estimator created by the Bilby samples. The vertical axis represents how many samples Bilby has generated (live and dead samples).}
    \label{fig:bilby_likelihood}
\end{figure}

\end{document}